\documentstyle[aps,preprint]{revtex}
\begin{document}
\draft
\title{\large 
Dynamical Localization: Hydrogen Atoms in Magnetic and Microwave fields
}
\author{Francesco Benvenuto, Giulio Casati\cite{infm}}
\address{Dipartimento di Fisica dell'Universit\`a, Via Castelnuovo 7, 22100
Como, Italy }
\author{Dima L. Shepelyansky\cite{binp}}
\address{Laboratoire de Physique Quantique, UMR C5626 du CNRS, \\
Universit\'e Paul Sabatier, 31062,
Toulouse, France}
\date{11 September 1996}
\maketitle
\begin{abstract}
We show that dynamical localization for
excited hydrogen atoms in magnetic and microwave
fields takes place at quite low microwave frequency 
\hbox{($\omega n^3 \ll 1$)}.
The  estimates of localization length are given for different parameter
regimes, showing that the
quantum delocalization border 
drops significantly as compared to the case of zero magnetic field. 
This opens up broad possibilities for laboratory investigations.
\end{abstract}

Quantum dynamical localization is one of the most interesting phenomena within
 the so--called field of
``Quantum Chaos'' \cite{ref1} .
This phenomenon, which consists in the quantum suppression of classically 
chaotic diffusive process, was discovered long ago \cite{ref2}, and has 
also been observed in several laboratory experiments 
\cite{ref3,ref4,ref5,ref6,ref7}. Even though experimental results
are in satisfactory agreement with theoretical predictions, the range of 
parameters over which quantum localization has been observed is still quite 
narrow. 
For example, in
experiments performed on hydrogen and Rydberg atoms in microwave fields
\cite{ref3,ref4,ref5}, in order to observe localization of the quantum
probability distribution around the initially excited state $n_0$, 
it is necessary that the localization length
$\ell_\phi$ of the distribution, measured in number of photons, be smaller
than the number $N_I$ of photons required for ionization.
Such a condition imposes severe experimental restrictions.
Indeed, $N_I = \left(2 n_{0}^{2} \omega\right)^{-1} = {n_0}/{2 \omega_{0}}$, 
where $\omega$ is the microwave field frequency and $\omega_0 = \omega 
n_{0}^{3}$ is the rescaled frequency; therefore, in order to increase $N_I$ it 
is either necessary to increase the initially excited principal quantum number 
$n_0$, or to decrease the rescaled frequency $\omega_0$.
On the other hand, in real experiments $n_0 \alt 100$ and the additional 
condition $\omega_0 > 1$ must be verified if the system is to be found
within the classically diffusive region (for sufficiently high field intensity
$\epsilon$). In practice one typically obtains $N_I \approx 20$.

An interesting possibility for increasing $N_I$, thus allowing a better 
numerical and experimental analysis of dynamical localization, would be the
consideration of a hydrogen atom within superimposed magnetic and microwave 
fields. Indeed, such a model, for a sufficiently strong magnetic
field, displays 
classical chaotic dynamics even in the absence of the microwave field 
\cite{ref8,ref9}.
When the microwave field is turned on, this pre--existing chaos immediately
leads to the onset of classical diffusive energy excitation, even when the
microwave field intensity $\epsilon$ goes to zero. Such excitation  takes
place even when $\omega_0 < 1$, therefore the quantity $N_I ={n_0}/{2
\omega_{0}}$ can be significantly increased.

Quantum effects may lead to localization of classical diffusion, hence
the 
critical microwave field intensity required for ionization or the so--called
quantum ionization border, will be determined by the condition $\ell_\phi
\ge N_I$. Indeed, only when the localization length $\ell_\phi$, expressed in
number of photons, is larger than the number $N_I$ of photons required for
ionization, will a non negligible amount of quantum probability  fall into the 
continuum states, thus leading to ionization.

In order to analyze the ionization process we consider the simple case in which
both fields are parallel. Due to the cylindrical symmetry, the magnetic
quantum number $m$ is an exact integral of the motion, and the Hamiltonian has 
the form:
\begin{equation}
H = {{p_z^2}\over{2}}+{{p_\rho^2}\over{2}}+{{m^2}\over{2\rho^2}}+
{{\omega_L m}\over {2}} + 
{{\omega_L^2 \rho^2}\over{8}}-{1\over\sqrt{z^2+\rho^2}}+
\epsilon z \cos \omega t
\label{eq1}
\end{equation}
where $\omega_L = B/c = B\hbox{\it(Tesla)}/B_0$ is the
rotation frequency in a magnetic field, 
$B_0 = 2.35 \times 10^5 \hbox{\it{Tesla}}$, 
$\epsilon$ and $\omega$ are the field 
intensity and frequency respectively (atomic units are used).

In the presence of a magnetic field only ($\epsilon=0$), according to well--known
results \cite{ref8,ref9}, the motion becomes strongly chaotic when $\omega_L 
n_0^3 \agt 4$ and no islands of stability are observed for $\omega_L n_0^3 
\approx 9$.
When the microwave field is turned on, diffusive excitation of the electron 
will take place and
the energy diffusion rate $D_B$ per unit time can be estimated in the following
way:
from  eq. (\ref{eq1}), $\dot E = - \epsilon z \omega \sin \omega t$ and 
therefore, for $\omega_0 \ll 1$ in a quasilinear approximation \cite{ref10},
the diffusion is determined by the spectral density of perturbation. That is,
$ D_B \sim \left(\epsilon z \omega\right)^2/\delta \omega \sim 
\left(\epsilon z \omega\right)^2 n_0^3 \sim D_0 \omega_0^2$, where
$\delta \omega \simeq n_0^{-3}$ is a typical frequency at $\epsilon = 0$ and
$D_0 = \epsilon^2 n_0 /2$ is the diffusion rate in the chaotic regime for $B=0$
and $\omega_0=1$ \cite{ref11}.
Therefore:
\begin{equation}
D_B = \chi_1 D_0 \omega_0^2
\label{eq2}
\end{equation}
where $\chi_1$ is a constant to be numerically determined.
This estimate is a good approximation if
$\omega_L n_0^3$ is not too large (the case $\omega_L n_0^3 \gg 10$ should be 
separately studied).

In the regime $\omega_0 \gg 1$ the energy change will take place only when the
electron passes close to the nucleus and, in analogy with the case $B=0$,
the diffusion rate is given approximately  by \cite{ref11} $D_B \simeq
\chi_2 \epsilon^2 n_0 \omega_0^{-4/3}/2 = \chi_2 D_0 \omega_0^{-4//3}$.

Notice that $D_B$ does not explicitly depend on the magnetic field $B$; however,
the quantities $\chi_1, \chi_2$ can both weakly depend on $B$ for
$\omega_{L} n_0^3 \agt 2$ (see fig~\ref{fig3}).

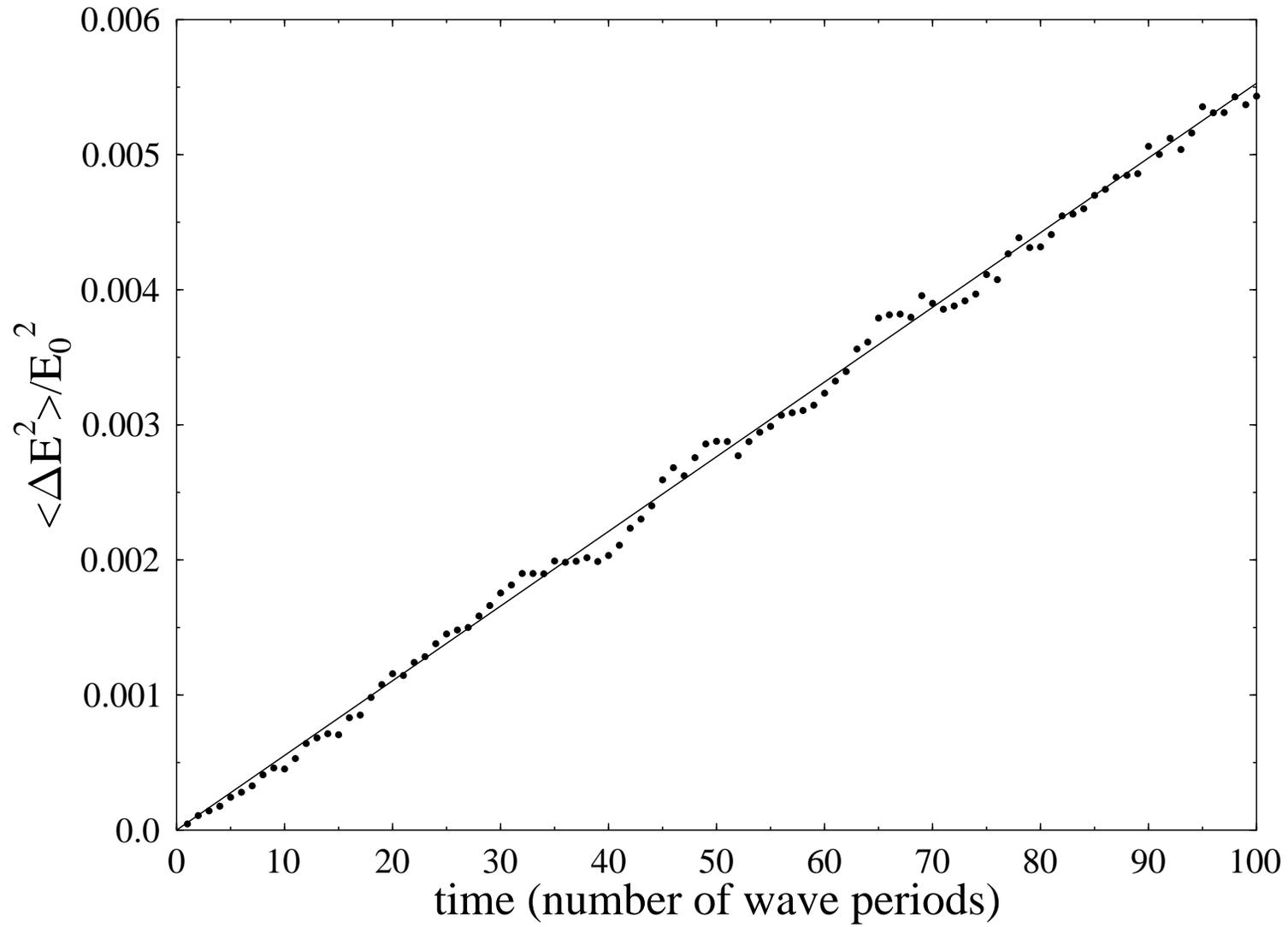
\begin{figure}
\caption{
Example of 
diffusive energy excitation 
$\langle\Delta E^2\rangle/E_0^2$ with time, for parameter values
$\omega_L n_0^3 = 9$,
$\omega_0=0.05$ and $\epsilon_0=0.003$. 
The straight line shows the diffusive growth with a fitted rate given by 
eq.(\protect\ref{eq2}) with $\chi_1 = 9.8$.}
\label{fig1}
\end{figure}

In order to check the above analytical estimates for $D_B$, we
have carried out
numerical simulations of the classical problem (\ref{eq1}). We fixed the
magnetic number $m=0$ and measured the energy excitation for an ensemble of
100 trajectories with the same initial energy $E_0 = -1/{(2 n_0^2)}$.
In fig.~\ref{fig1} we demonstrate the diffusive excitation process in energy for
scaled frequency $\omega_0 \ll 1$ and very small microwave field $\epsilon_0 =
\epsilon n_0^4$. Here $\omega_L n_0^3 = 9$ so that the motion is 
fully chaotic even for $\epsilon_0=0$.

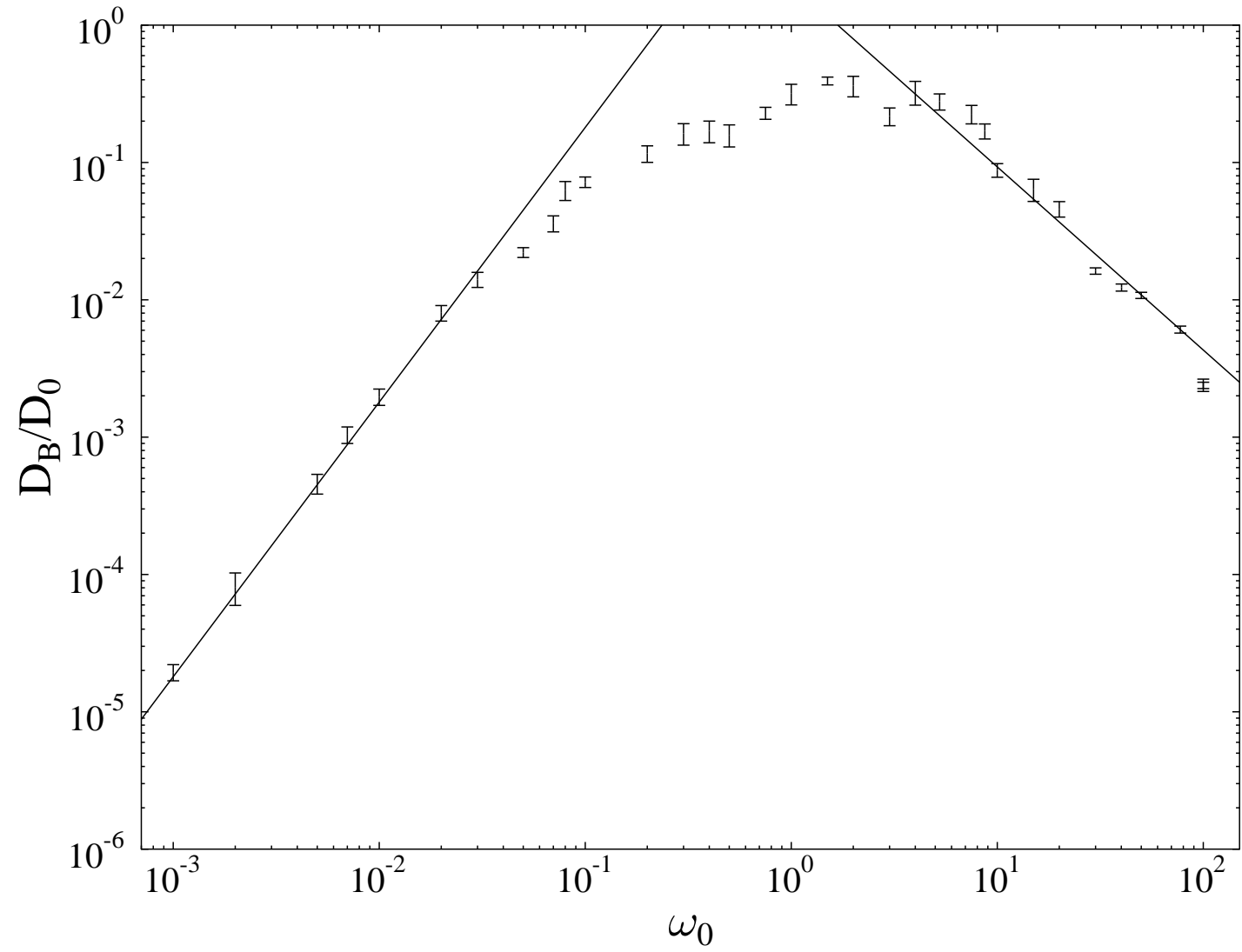
\begin{figure}
\caption{
Dependence of
scaled diffusion rate $D_B/D_0$
on scaled frequency $\omega_0$ for $\omega_L n_0^3 =9$. Here 
$D_0 = \epsilon_0^2/2 n_0^7$ and 
$\epsilon_0$ in the interval $10^{-3} \le \epsilon_0 \le 10^{-2}$.
The straight lines show the theoretical dependence for $\omega_0 \gg 1$
and $\omega_0 \gg 1$ with $\chi_1 = 18$ and $\chi_2 = 2$ respectively 
(see text).
Ensembles from 30 to 100 orbits have been used.
}
\label{fig2}
\end{figure}

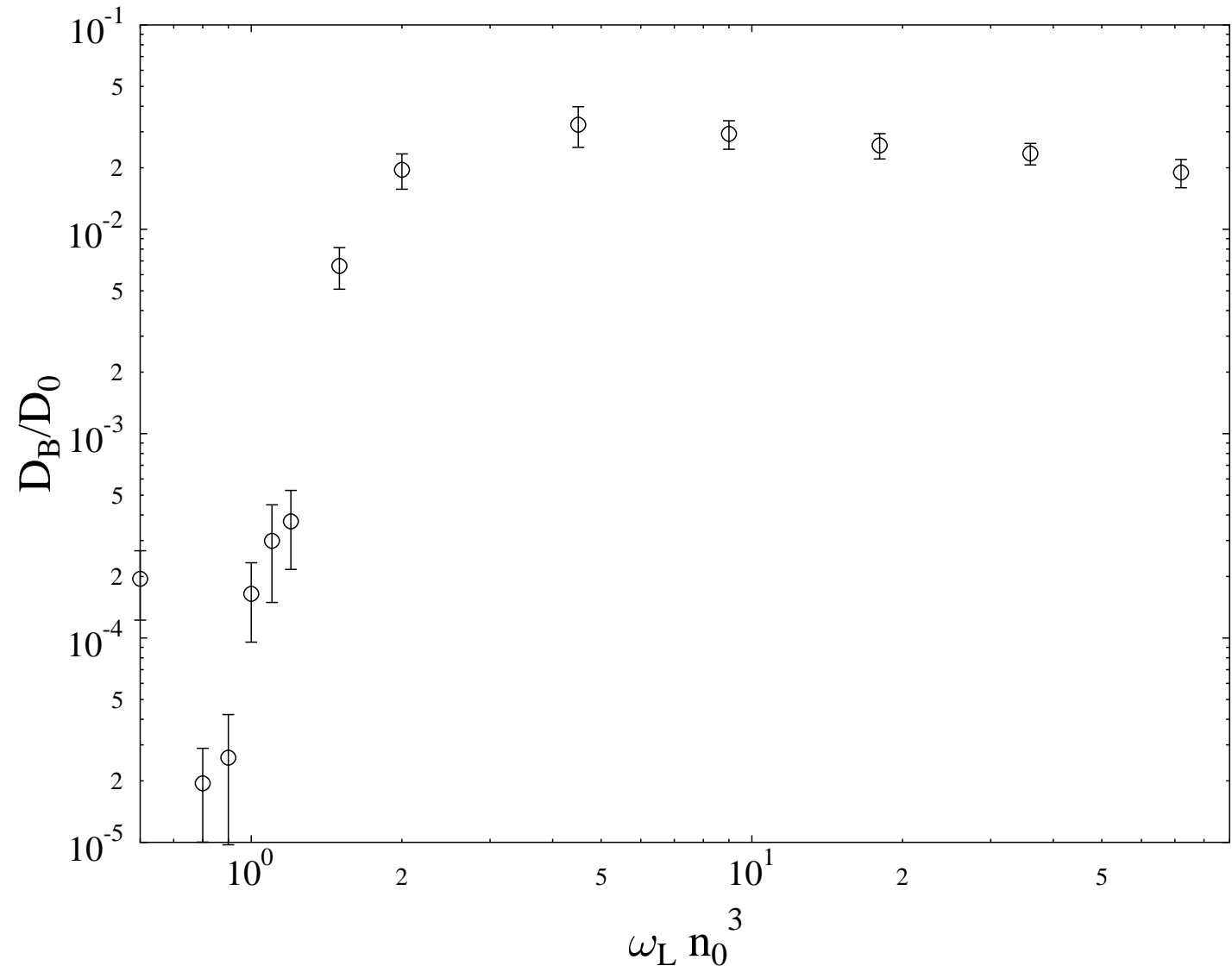
\begin{figure}
\caption{Scaled diffusion rate $D_B/D_0$
as a function of the scaled frequency $\omega_L n_0^3$ for
$\omega_0 = 0.05$
and $\epsilon_0=0.003$.
}
\label{fig3}
\end{figure}

The frequency dependence of the diffusion rate $D_B$  is shown in
fig.~\ref{fig2} for fixed $\omega_L n_0^3  = 9$. The behavior for small and
large frequencies is in agreement with the above theoretical estimates and
is indicated by the two straight lines with $\chi_1 \approx 18$ and 
$\chi_2 \approx 2$ respectively. An interesting feature is that in a rather
large frequency interval ($10^{-1} \alt \omega_0 \alt 10$) the frequency
dependence is rather weak.
We also checked the $\epsilon^2$ dependence of the diffusion rate $D_B$ by
varying the field intensity over a few orders of magnitude.

In fig.~\ref{fig3} we plot the dependence of $D_B$ on magnetic field intensity
for small values of $\omega_0$. According to our expectations, the diffusion
rate remains approximately constant for $\omega_L n_0^3 > 2$ while below
this value the internal motion becomes integrable and diffusion drops to zero.

The classical diffusive process will lead to ionization of the atom after a
time $t_I \approx E_0^2/D_B$. For low field strength this time is much larger
than both the orbital period and the microwave field period, and therefore
quantum interference effects can lead to localization of classical diffusion.
The localization length $\ell_\phi$ (in number of photons) can be obtained
by using known estimates \cite{ref12}, according to which
$\ell_\phi \sim \Gamma\rho$, where  $\rho$ is the density of coupled
states and  $\Gamma$ is the one--photon transition rate, which
in turn can be
related to the classical diffusion rate via $\Gamma=D_B \omega^{-2}$.
The striking advantage of such an approach to the computation of
the localization length
\cite{ref12} is that it involves only 
classical characteristics of motion, namely
the classical diffusion rate and density of coupled states.
 It is therefore sufficient to know only these two quantities to understand 
the properties of quantum dynamics. In this way one need not worry about
quantum matrix elements, the structure of eigenfunctions and other quantum
 aspects.
This fact becomes even more remarkable if one remembers that
without a magnetic field the effective dynamics is well described
by a one-dimensional atom model \cite{ref11} while in a strong magnetic
field the motion is essentially two-dimensional and quite different
from the usual Kepler dynamics.

For zero magnetic field the effective density of coupled states is $\rho = n^3$,
due to Coulomb degeneracy and the existence of an additional approximate
integral of motion
 \cite{ref11}. Instead, for sufficiently large $B$, the internal motion is 
chaotic and the density of coupled states is $\rho = n^4$. 
Therefore,
in the latter case the localization length in number of photons is:
\begin{equation}
\ell_B = \ell_\phi {{D_B}\over{D_0}} {{n_0}\over{\omega_0^2}}
\label{eq3}
\end{equation}
where $\ell_\phi= 3.3 \epsilon_0^2 n_0^2$ is the localization length 
in the microwave field only ($B=0$) at $\omega_0 = 1$ and it is assumed
that $l_B >1$.

For $\omega_0 \ll 1$ we obtain from (\ref{eq3}) $\ell_B = 3.3 \chi_1
 \epsilon_0^2 n_0^3$. The delocalization border, namely the critical 
microwave field intensity $\epsilon_1$ above which ionization takes
place, is given by the condition $\ell_B = N_I$ leading to:
\begin{equation}
\epsilon_{01} = {1\over{n_0\sqrt{6.6\chi_1 \omega_0}}}
\label{eq4}
\end{equation}
which is well below the delocalization border within the microwave field
alone.
We would like to stress that the above estimates are valid in the regime 
$\omega \rho > 1$ \cite{ref12} that is $\omega_0 > n_0^{-1}$ (and therefore
the border (\ref{eq4}) is valid for $ n_0^{-1} < \omega_0 < 1$).

The above estimates were made in analogy to the problem
of photonic localization in a random
molecular quasi-continuum \cite{ref12}. However, it is possible 
that such approach should be modified for 
the present case of a well defined quasiclassical system
with adiabatically slow perturbation. Indeed, in such a situation,
one can make a transformation to an instantaneous time basis
in which transition matrix elements are proportional to the frequency of 
the perturbation. Here, many levels will be mixed only if this
effective matrix element $V_{eff}$ proportional to $\omega$ is larger
than the level spacing in one atomic shell $n^{-4}$.
For a chaotic case we can estimate 
$V_{eff} \sim \epsilon \omega n^5/\sqrt{n}$ so that
the diffusive exitation is quasiclassical (nonperturbative)
only if $V_{eff} > 1/n^4$. The same estimate can be also understood
in the following quasiclassical way. The diffusion rate in the 
number of shells (the change of the principal quantum number
in the unperturbed hydrogen $\Delta n =1$) is $D_n = D_B n^6$.
All the levels in the shell will be resolved after time
$t_{sh} \approx 2 \pi n^4$. Therefore, the adiabatic perturbation
is perturbative if the number of shells mixed during this time is less than 1:
$(\Delta n)^2 \approx D_n t_{sh} <1$. 
The situation is quasiclassical in the opposite case 
$(\Delta n)^2 > 1$ when:
\begin{equation}
\epsilon_{0} > \epsilon_{0 a} \approx ({{1 }\over {3 \chi_1}})^{1/2} 
{{1} \over {\omega_0 {n_0}^{3/2}}}
\label{eq4a}
\end{equation}
In this regime the diffusion rate measured in the number of shells
per unit of time $\tau$ measured in units of $t_{sh}$ is 
$D_{sh} = (\Delta n)^2/{\Delta \tau} 
\approx 3 \chi_1 {\epsilon_0}^2 {\omega_0}^2 n_0^3 > 1$.
Similarly to the standard
estimate \cite{ref12} this diffusion will be localized after 
time $\tau^* \approx D_{sh}$ with localization length
$l_n \approx \Delta n (\tau^*) \approx D_{sh}$
(in the number of photons this gives $l_B=l_n/\omega_0=D_{sh}/\omega_0$).
Then the delocalization border is determined by the condition
$l_n \sim n_0/2$ giving
\begin{equation}
\epsilon_{0} > \epsilon_{0qa} \approx {1 \over {\sqrt{6 \chi_1} \omega_0 n_0}}
\label{eq4b}
\end{equation}
As we see these adiabatic estimates are different from the above eqs. (3), (4).
In our opinion the physical situation is more correctly described
by the approach (3) - (4). However, only direct quantum simulations 
or laboratory experiments can lead to more definite conclusions.

For $\omega_0 \gg 1$ the situation is more standard and we have:
\begin{equation}
\ell_B = 3.3 \chi_2 {{\epsilon^2}\over{\omega^{10/3}}} n_0
\label{eq5}
\end{equation}
and the condition $\ell_B = N_I$ leads to:
\begin{equation}
\epsilon_{02} = {{\omega_0^{7/6}}\over{n_0\sqrt{6.6\chi_2}}}
\label{eq6}
\end{equation}
which is $\sqrt{n_0 \chi_2}$ times smaller than the quantum
delocalization border for $B=0$.

It is important to remark that, contrary to the case $B=0$, the
localization length depends now on photon number. Indeed, for $\omega_0
\ll 1$, $\ell_B \propto n_0^{11}\sim \left(N_I-N\right)^{-11/2}$ while for
$\omega_0 >1$, $\ell_B \propto n_0 \sim \left(N_I-N\right)^{-1/2}$.
This means that localization length becomes larger near the ionization border.
However, this effect is not important if $\ell_B \ll N_I$ while for
$\ell_B \approx N_I$ it will lead to a change in the numerical factor 
for the delocalization border.

We also remark that for $\omega_0 \ll 1$ the quantum delocalization 
border $\epsilon_{01}$ will be relevant only if it remains 
smaller than the ionization border for a static electric field in the
presence of a parallel magnetic field.

Finally, we would like to stress that the problem under investigation represents
a qualitatively new situation as compared to the hydrogen atom in a 
microwave field only. Therefore, the above preliminary estimates
require a detailed check by quantum simulations.

On  qualitative grounds, the effects described 
in the present paper can also take place
 in non hydrogen atoms in a static electric field and 
in the absence of a magnetic 
field. It is indeed known that for non hydrogen atoms in static electric
fields, the eigenfunctions can be approximately ergodic and the level spacing
statistics shows level repulsion of the Wigner-Dyson type \cite{ref13}.
This situation is similar to that in which internal chaos is present
due to a strong magnetic field. Since in the formulae derived in this paper
the magnetic field dependence is quite weak, we expect that they are
approximately valid even for non hydrogen atoms in static electric fields.
However, one should keep in mind that while qualitatively
the ionization process should behave  similarly to the case with
a magnetic field
the chaos properties in a static electric field are not so strong
and therefore this case should be analyzed in more detail.
Such a situation with atoms in a static electric field is quite interesting
since it is more suitable for laboratory experiments.
Finally, we believe that the results described in the present paper will not be
significantly changed if the electric and magnetic fields are not 
parallel, provided that the electric field is relatively weak. The reason
being that the magnetic quantum number $m$ will determine the existence
of an approximate integral of motion similar to the case of
hydrogen in a microwave field \cite{ref11}.

In this paper we have introduced and discussed a new model that should 
allow for a clear experimental investigation of quantum dynamical localization.
Its main feature is the wide range inside which it is possible to change
physical parameters such as $\omega_0$, $\epsilon_0$ and the interaction
time. A possible convenient choice of parameters is $n_0 = 60$, 
$\omega_0 = 0.1$ ($\omega = 3. \hbox{GHz}$), $B=10 T$ ($\omega_L n_0^3 = 9 $)
[or static electric field $\epsilon_{0st} \approx 0.05$ for
non hydrogen atoms].
For these conditions we have
$\epsilon_{01} = 0.005$, $N_I=300$ and interaction time
(measured in number of microwave periods): $\tau_I =
\omega t_I / 2 \pi  = 1/(4 \pi \chi_1) \cdot 1/(\epsilon_0^2
\omega_0) \approx 1800$.
If, however, we define the ionization probability as the
occupation probability
 of levels
$n>70$ then the number of photons will be $N_I\sim 100$ and the ionization
time $\tau_I = 180$ microwave periods. This range of parameters seems to be
suitable for laboratory experiments.

\end{document}